\documentclass{elsart}
\usepackage{graphicx}
\usepackage{amssymb}
\begin{document}
\begin{frontmatter}
\title{ Critical revision of the ZEPLIN-I sensitivity to WIMP interactions }

\author{ A. Benoit$^{1}$,}
\author{ M. Chapellier$^{2}$,}
\author{ G. Chardin$^{3}$,}
\author{ L. Dumoulin$^{4}$,}
\author{ K. Eitel$^{5}$,}
\author{ J. Gascon$^{6}$,}
\author{ G. Gerbier$^{3}$,}
\author{ C. Goldbach$^{7}$,}
\author{ J. Jochum$^{8}$,}
\author{ A. de Lesquen$^{3}$,}
\author{ G. Nollez$^{7}$,}
\author{ F. Pr\"obst$^{9}$,}
\author{ W. Seidel$^{9}$}

\address{$^{1}$Centre de Recherche sur les Tres Basses Temperatures,
SPM-CNRS, BP 166, 38042 Grenoble, France}
\address{$^{2}$DSM/DRECAM, CEA/Saclay,\\
F-91191 Gif-sur-Yvette Cedex, France}
\address{$^{3}$DSM/DAPNIA, CEA/Saclay,\\
F-91191 Gif-sur-Yvette Cedex, France}
\address{$^{4}$Centre de Spectroscopie Nucleaire et de Spectroscopie de Masse,
      IN2P3-CNRS, Universite Paris XI,
      bat 108, 91405 Orsay, France }
\address{$^{5}$Forschungszentrum Karlsruhe, Institut f\"ur Kernphysik,\\
Postfach 3640, 76021 Karlsruhe, Germany}
\address{$^{6}$Institut de Physique Nucleaire de Lyon-UCBL, IN2P3-CNRS,
4 rue Enrico Fermi, 69622 Villeurbanne Cedex, France}
\address{$^{7}$Institut d'Astrophysique de Paris, INSU-CNRS, \\
98 bis Bd Arago, 75014 Paris, France}
\address{$^{8}$Physikalisches Institut - Universit\"at T\"ubingen,\\
Auf der Morgenstelle 14, 72076 T\"ubingen, Germany}
\address{$^{9}$Max-Planck-Institut f\"ur Physik, F\"ohringer Ring 6, \\
80805 M\"unchen, Germany}
\begin{abstract}
The ZEPLIN collaboration has recently published its first result presenting a maximum sensitivity of $1.1 \times 10^{-6}$ picobarn for a WIMP mass of $\approx$ 60 GeV. The analysis is based on a discrimination method using the different time distribution of scintillation light generated in electron recoil and nuclear recoil interactions. We show that the methodology followed both for the calibration of the ZEPLIN-I detector response and for the estimation of the discrimination power is not reliable enough to claim any background discrimination at the present stage. The ZEPLIN-I sensitivity appears then to be in the order of 10$^{-3}$ picobarn, three orders of magnitude above the claimed 1.1 10$^{-6}$ picobarn.
\end{abstract}
\end{frontmatter}

\section{Introduction }
Most of the progress in Dark Matter Direct Detection sensitivity is associated with the advent of a new generation of discriminating detectors, able to reject the important radioactive background associated mostly with electron recoils, and making it possible to identify a small population of nuclear recoil interactions. Three cryogenic experiments, CDMS~\cite{cdms}, EDELWEISS~\cite{edelweiss} and CRESST~\cite{cresst}, have applied this discrimination scheme using the simultaneous measurement of charge and phonon signals, and of light and phonon signals. These three experiments have published sensitivities to spin-independent WIMP interactions ranging from $\approx$~1.6 10$^{-7}$ picobarn to $\approx$~2.0 10$^{-6}$  picobarn at maximum sensitivity~\cite{cdms,edelweiss,cresst}.
The ZEPLIN-I experiment has followed a parallel discrimination strategy by measuring two quantities, the visible energy and the scintillation time constant. Despite a much higher background rate, an energy resolution of 100 \% or larger for events of interest and a statistical discrimination, this experiment claims a sensitivity of 1.1 10$^{-6}$ picobarn, requiring a background subtraction at the 99.9 \% level~\cite{zeplin}.

\section{Neutron calibrations}
For all discriminating experiments, background discrimination performances are determined by exposing the detector to gamma-ray and neutron sources and/or beams. The neutron calibrations of the charge-phonon and light-phonon detectors of the CDMS, EDELWEISS and CRESST experiments present nuclear recoil and electron recoil populations clearly separated, on an event-by-event basis, down to low energies ($\approx$~10 keV recoil energy).

The situation is markedly different for the ZEPLIN-I neutron calibration described in Ref. ~\cite{zeplin}. The limited energy resolution and separation of the scintillation time constants allows only a statistical discrimination. Figs.~[1-3], reproduced after this publication~\cite{zeplin}, present the scintillation time constant distribution for three data samples. For each of these data samples, we present the original figure of the ZEPLIN-I publication, together with the same figure redrawn in linear scale without the fit to the data for additional clarity.

It should be kept in mind that the Xe nuclear recoil energies of interest for dark matter search are mostly below 40 keV (real recoil energy) for a WIMP mass of 60 GeV. Also, a recoiling nucleus provides less light than an electron of the same kinetic energy. The ratio between these two quantities is called the quenching factor, which represents the relative scintillation efficiency between a nuclear recoil and an electron recoil of the same real recoil energy. This quenching factor varies between $\approx$ 0.13 at 10 keV real recoil energy and $\approx~0.18$ at 50 keV real recoil energy, the energy interval of interest~\cite{aprile}.

\subsection{AmBe neutron source present, low energy interval 
\protect\\([3-8] keV visible energy, [20-48] keV recoil energy interval)}
\begin{figure}[hbtp]
\includegraphics[width=.9\textwidth]{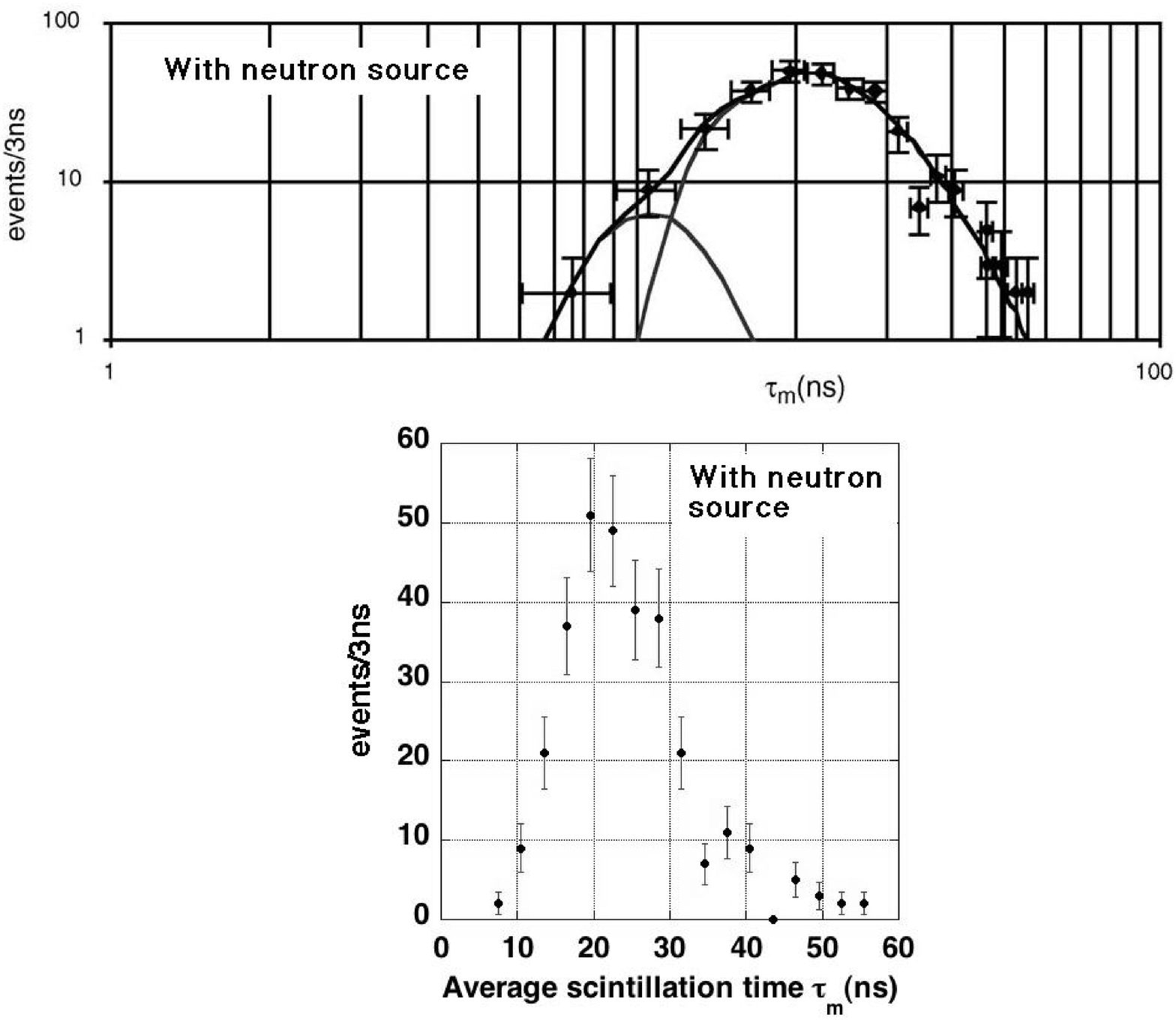}   
\caption{ Distribution of the average scintillation time constant for events recorded by ZEPLIN-I exposed to an Am-Be neutron source, visible energy range 3-8 keV, real recoil energy interval $\approx$ 20-48 keV (see text). It is highly questionable to attribute about 8 events in the two left bins to a nuclear recoil population. Upper part: original figure after Ref.~\cite{zeplin}, lower part: same figure redrawn in linear scale.\label{zeplin_fig2}}
\end{figure}

Figure~\ref{zeplin_fig2} has been obtained by exposing the detector to an Am-Be neutron source in a lab at the earth surface. No neutron calibration has been performed in the underground site. This figure presents the scintillation time distribution for events with visible energy between 3 and 8 keV. This corresponds to real recoil energies between $\approx$ 20 keV and $\approx$ 48 keV~\cite{aprile}. Although most nuclear recoils on xenon are expected to lie in this lower energy interval, no visible excess of events can be seen on the figure. This is particularly clear when the figure is drawn in linear scale. It should also be noted that a noise cut is applied on all events with scintillation time constant $<$ 10 ns and with more than 2.5 photoelectrons in any photomultiplier. The tentative nuclear recoil "peak" around 10 ns is then dangerously close to photomultiplier noise and strongly influenced by the noise cut applied on the data. It is therefore not legitimate to attribute to nuclear recoils the $< 10$ events at scintillation time constant $\approx$ 10 ns.

This small nuclear recoil population strongly contrasts with the recoil distribution spectrum of the neutron calibrations of all other discriminating experiments, where indeed nuclear recoils dominate over electron events at low energies.

\subsection{AmBe neutron calibration source present, high energy interval 
\protect\\([20-30] keV visible energy, [100-150] keV recoil energy interval)}
\begin{figure}[hbtp]
\includegraphics[width=.9\textwidth]{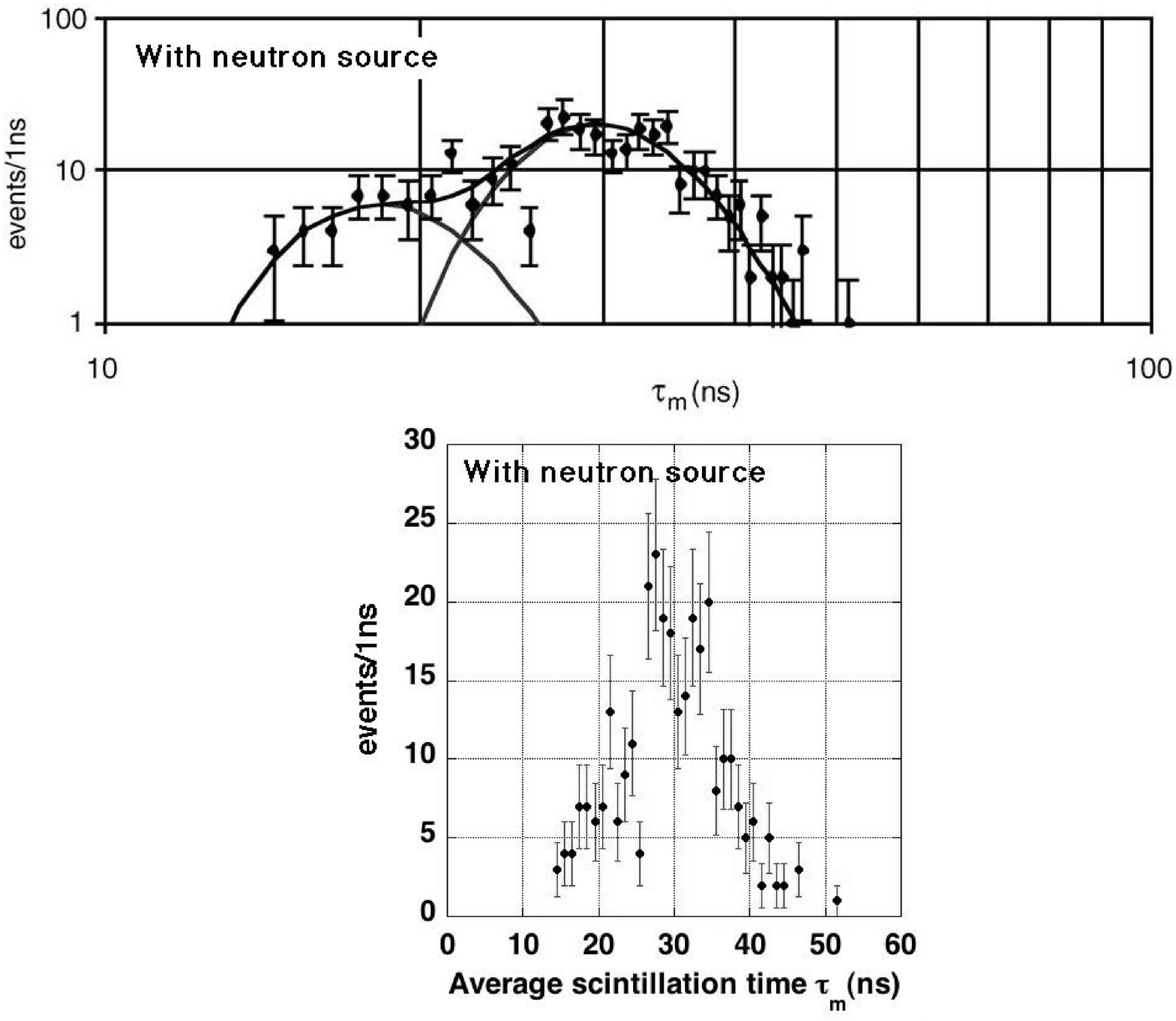}   
\caption{ Distribution of the average scintillation time constant for events recorded by ZEPLIN-I exposed to an Am-Be neutron source, visible energy range 20-30 keV, real recoil energy interval $\approx$ 100-150 keV. Upper part: original figure after Ref.~\cite{zeplin}, lower part: same figure redrawn in linear scale.\label{zeplin_fig1}}
\end{figure}

Figure~\ref{zeplin_fig1} has also been obtained by exposing the detector to an Am-Be neutron source. The scintillation time distribution for events with visible energy between 20 and 30 keV is represented. Converted in terms of real recoil energy, the energy interval is then $\approx$ 100 to 150 keV \cite{aprile}. 

About twenty events are observed in the region attributed to nuclear recoils, about one order of magnitude less than the number of gammas recorded in the same energy interval. The average scintillation time constant of nuclear recoils is of the order of $\approx$ 20 ns. Although very small, the nuclear/gamma signal ratio is larger in this high energy interval than in the low energy interval, in strong contrast with the expected distribution and with data from the other discriminating experiments.

This inconsistency is not addressed in the ZEPLIN publication and no simulation is shown demonstrating that this peculiar feature is expected.

\subsection{Neutron source removed, low energy interval
\protect\\([3-7] keV visible energy, [20-42] keV recoil energy interval)}
\begin{figure}[hbtp]
\includegraphics[width=.9\textwidth]{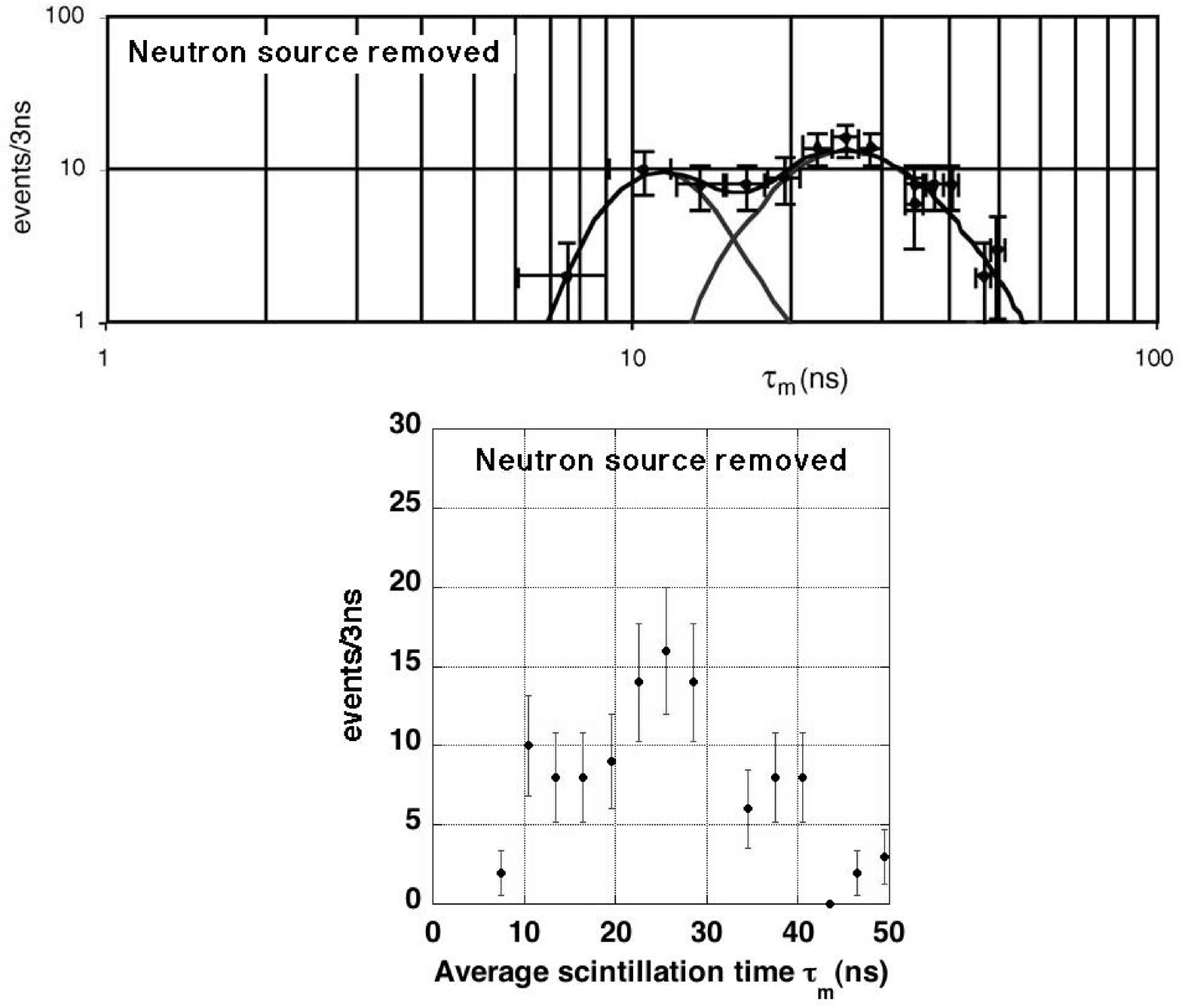}   
\caption{ Distribution of the average scintillation time constant for events recorded by ZEPLIN-I {\it with the Am-Be neutron source removed}, visible energy range 3-7 keV, real recoil energy interval $\approx$ 20-42 keV (see text). Surprisingly, this data sample has the best signal/noise ratio of the three data samples presented by ZEPLIN-I, with about 30 "ambient neutron" candidate events at scintillation time constants 10-15 ns. Upper part: original figure after Ref.~\cite{zeplin}, lower part: same figure redrawn in linear scale. \label{zeplin_fig3}}
\end{figure}

Figure~\ref{zeplin_fig3} presents the scintillation time distribution for events with visible energy between 3 and 7 keV. This corresponds to real recoil energies between $\approx$ 20 keV and $\approx$ 42 keV~\cite{aprile}. This data sample presents the best neutron/gamma signal ratio among the three neutron calibration data samples presented by ZEPLIN-I.

But, surprisingly, this data sample has been obtained by {\it removing the neutron source}. The events presenting scintillation time constants in the 10-15 ns time interval are then attributed to recoils induced by "ambient neutrons". The number of these events is $\leq$ 30 and the average time constant of these "nuclear recoil" events is significantly different from the nuclear recoil candidate events of Figure~\ref{zeplin_fig1} ($\approx$~11 ns compared to $\approx$ 20 ns). Here also, a noise cut is applied on the data to all events with scintillation time constant $<$ 10 ns and with more than 2.5 photoelectrons in any photomultiplier. There is no demonstration that the "ambient neutron" sample is not due to photomultiplier noise or to cosmic-ray induced events (since all neutron calibrations were done at surface). 

\begin{figure}[hbtp]
\includegraphics[width=.9\textwidth]{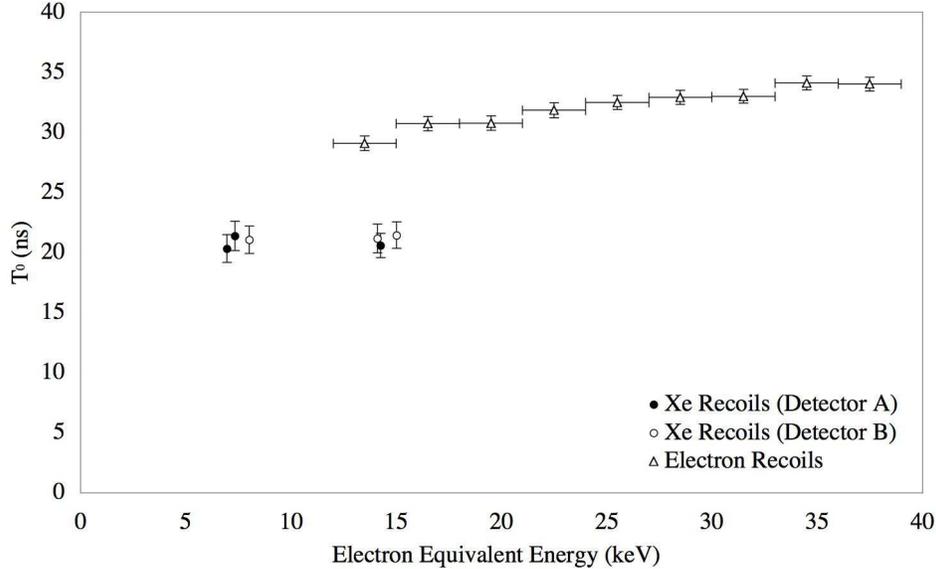}   
\caption{Average scintillation time constant recorded for electron and nuclear recoil interactions, after Ref.\protect~\cite{akimov}. In this calibration data by the same group, contrasting with the ZEPLIN-I analysis, nuclear recoil time constants appear constant down to 7 keV visible energy ($\approx$ 42 keV real recoil energy), while the average scintillation time constant of electron recoils decreases with energy. \label{akimov_fig}}
\end{figure}

No explanation is given of the sudden drop of the "nuclear recoil" time constant from 20 ns to 10 ns at the lowest energies. Calibration data previously recorded by the same collaboration indicate just the opposite trend: nuclear and electron recoil scintillation time constants appear to converge when the detected energy is decreased (see Figure~\ref{akimov_fig}, reproduced after Ref. ~\cite{akimov}). Also, NaI(Tl) and CsI(Tl) scintillators show similar trends of vanishing discrimination at the lowest energies~\cite{gerbier,pecourt}.

No explanation is given as to why it is by {\it removing the neutron source} that the highest neutron/gamma signal ratio is obtained by ZEPLIN-I. No other group has confirmed this discrimination method.

We conclude that there is at present no demonstration that nuclear recoil events have scintillation time constants different from electron recoil events at low energies. On the contrary, there are several indications that discrimination vanishes at these energies. Therefore, the discrimination procedure and the background subtraction technique applied by ZEPLIN appear invalid. All events must then be considered as potential WIMP candidates, and the resulting sensitivity is then about 10$^{-3}$ picobarn.

\section{Conclusions}

No demonstration has been given in the ZEPLIN-I analysis that nuclear recoils and electron recoils differ at the low energies (a few keV of visible energy) where most WIMP interactions are expected. Previous calibrations by the same group indicate just the opposite: nuclear and electron recoil scintillation time constants appear to converge when the detected energy is decreased.

In addition, the calibration procedure described in the latest ZEPLIN-I publication shows a major methodological problem: it is when the neutron source is removed that the highest neutron/gamma signal ratio is observed. A tentative but unproven population of "ambient neutrons", in the absence of a neutron source, is then used as a calibration, but a noise cut overlaps the "ambient neutron" selection. Therefore, it is not possible to exclude at present that this small "ambient neutron" population is in fact a photomultiplier noise tail, induced for instance by cosmic rays.

Waiting for more precise and consistent measurements, in an energy region where the present resolution is everywhere $\geq 100\%$, all events recorded at low energies should be considered as potential WIMP candidates. The maximum ZEPLIN-I sensitivity then becomes conservatively 10$^{-3}$ picobarn, three orders of magnitude above the claimed 1.1 10$^{-6}$ picobarn.

\end{document}